\documentclass[aps,prl,twocolumn,groupedaddress,showpacs]{revtex4}

\usepackage{graphicx}
\begin{document}


\title{Spin-parity dependent tunneling of magnetization 
in single-molecule magnets}



\author{W. Wernsdorfer$^1$, S. Bhaduri$^2$, C. Boskovic$^2$, 
G. Christou$^2$, D.N. Hendrickson$^3$}

\affiliation{
$^1$Lab. L. N\'eel, associ\'e \`a l'UJF, CNRS, BP 166,
38042 Grenoble Cedex 9, France\\
$^2$Dept. of Chemistry and Molecular Structure Center, Indiana Uni., 
Bloomington, 47405-4001, IN, USA\\
$^3$Dept. of Chemistry and Biochemistry, Uni. of California at San 
Diego, La Jolla, 92037, CA, USA
}


\date{\today}

\begin{abstract}
Single-molecule magnets facilitate the study of quantum tunneling of 
magnetization at the mesoscopic level. The spin-parity effect is 
among the fundamental predictions that have yet to be clearly observed. It 
is predicted that quantum tunneling is suppressed at zero 
transverse field if the total spin of the magnetic system is 
half-integer (Kramers degeneracy) but is allowed in integer 
spin systems. The Landau-Zener method is used to measure the 
tunnel splitting as a function of transverse field. Spin-parity 
dependent tunneling is established by comparing the 
transverse field dependence of the tunnel splitting of integer and 
half-integer spin systems. 
\end{abstract}


\maketitle

Single-molecule magnets (SMM) are considered the best 
systems for studying quantum tunneling of magnetization at the mesoscopic 
level. The first molecule shown to be a SMM 
was Mn$_{12}$ acetate~\cite{Sessoli93}. 
It exhibits slow magnetization relaxation of its $S$ = 10 ground state 
which is split by axial zero-field splitting. It was the first system 
to show thermally assisted tunneling of magnetization
~\cite{Novak95,Friedman96,Thomas96}. 
Fe$_8$ and Mn$_4$ SMMs were the first to exhibit ground state 
tunneling~\cite{Sangregorio97,Aubin98}. 
Tunneling was also found 
in other SMMs (see, for instance,~\cite{Caneschi99,Price99,Yoo_Jae00}). 
A detailed study of the influence of 
environmental degrees of freedom on the tunnel process has been 
started on Fe$_8$ and Mn$_{12}$ acetate (concerning 
phonons~\cite{Kent00b,WW_EPL00}, 
nuclear spins and dipolar couplings~\cite{WW_PRL99,WW_EPL99,WW_PRL00}) 
which were motivated by theories 
\cite{Fort98,Chudnovsky97,Prokofev98}. 
The spin-parity effect is among the 
fundamental predictions which have yet to be established at the 
mesoscopic level. It is predicted that quantum tunneling is 
suppressed at zero applied field if the total spin of the magnetic 
system is half-integer but is allowed in integer spin systems. 
Van Hemmen and S$\ddot{\rm u}$to~\cite{VanHemmen86} 
were the first to suggest the 
absence of tunneling as a consequence of Kramers degeneracy
~\footnote{
The Kramers theorem asserts that no matter how unsymmetric the 
crystal field, an ion possessing an odd number of electrons must have 
a ground state that is at least doubly degenerate, even in the 
presence of crystal fields and spin-orbit interactions [H. A. 
Kramers, Proc. Acad. Sci. Amsterdam {\bf 33}, 959 (1930)]
The Kramers theorem can be found in standard textbooks on quantum 
mechanics L. D. Landau and E. M. Lifschitz, Quantum Mechanics 
(Pergamon, London, 1959).}. 
It was then shown that tunneling can even be absent without 
Kramers degeneracy~\cite{Loss92,Delft92,Garg93}. 
In this case, quantum phase interference can lead to 
destructive interference and thus 
suppress tunneling~\cite{Garg93}. This 
effect was recently seen in Fe$_8$ 
and Mn$_{12}$ SMMs~\cite{WW_Science99,WW_JAP02}.

There are several reasons why the first 
attempts~\cite{Aubin98,Gomes00} to observe 
the spin-parity effect were unsuccessful. The main reason reflects 
the influence of environmental degrees of freedom that can induce or 
suppress tunneling: Firstly, hyperfine and dipolar couplings can 
induce tunneling via transverse field components; furthermore, 
intermolecular superexchange coupling may enhance or suppress tunneling 
depending on its strength; phonons can induce transitions via excited 
states; and finally, faster relaxing species can complicate the 
interpretation~\cite{WW_EPL99}. 

In this letter, we show that these problems can be overcome by 
studying the tunnel splitting as a function of transverse 
field. We selected three SMMs which revealed to be sufficiently small 
to show clearly ground state tunneling. 

\begin{figure}
\includegraphics[width=.42\textwidth]{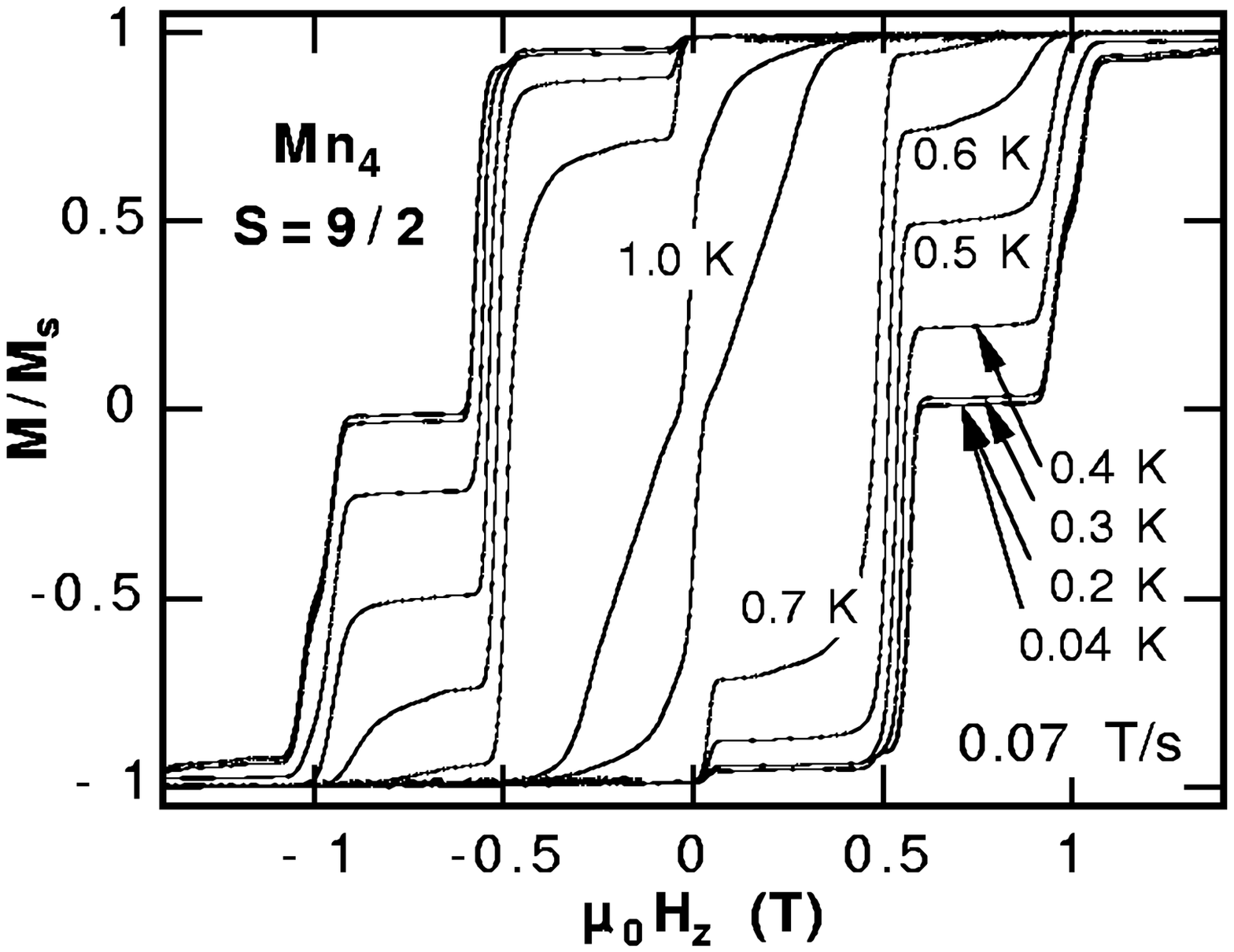}
\includegraphics[width=.42\textwidth]{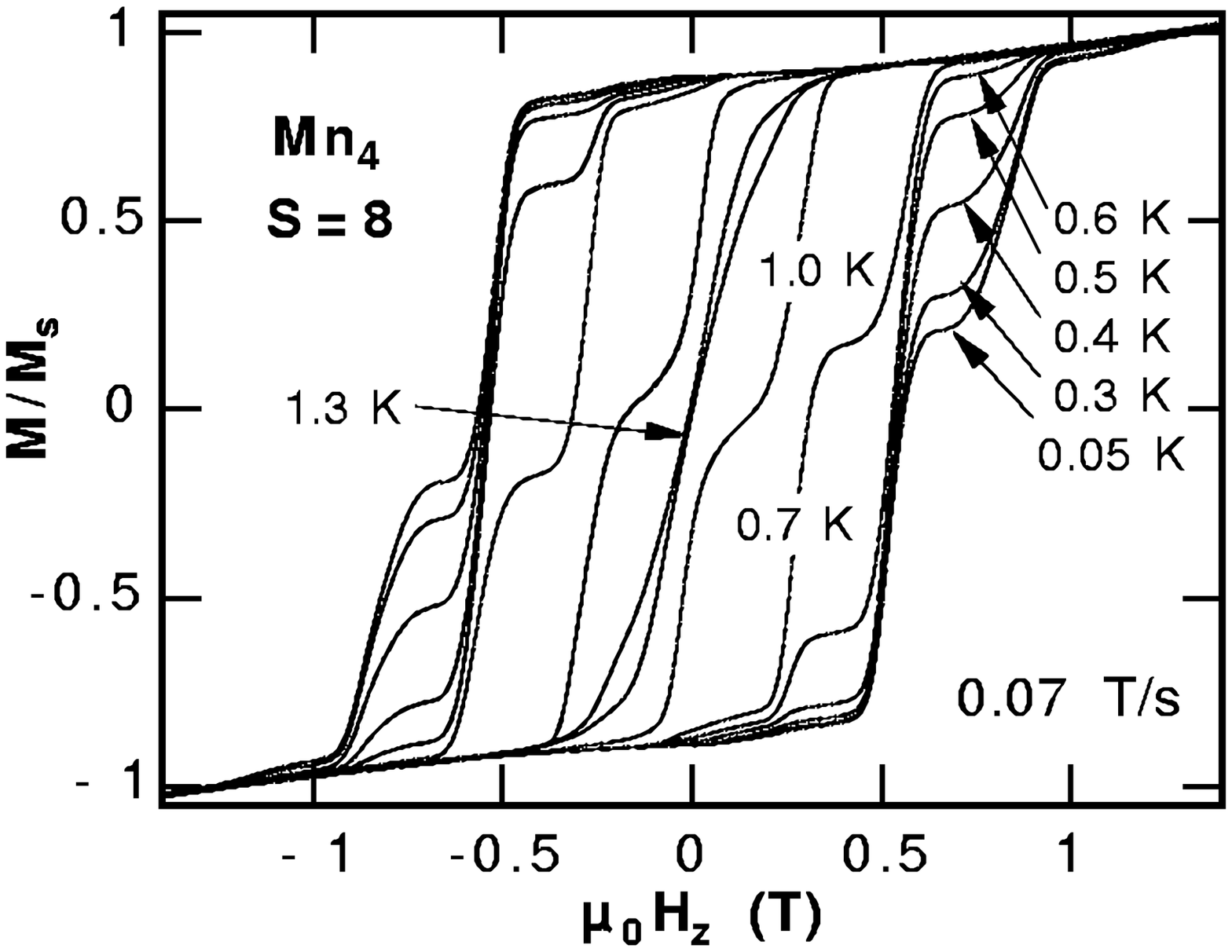}
\caption{Hysteresis loops of a single crystal 
of (a) Mn$_4$-(S=9/2) and (b) Mn$_4$-(S=8) molecular clusters
at different temperatures and a constant field
sweep rate $dH_z/dt$ = 0.07 T/s.}
\label{hyst}
\end{figure}

The first is [Mn$_4$O$_3$(OSiMe$_3$)(OAc)$_3$(dbm)$_3$], 
called Mn$_4$-($S=9/2$), with a 
half-integer ground state S = 9/2. 
The second is [Mn$_4$(O$_2$CMe)$_2$(Hpdm)$_6$][ClO$_4$]$_2$, 
called Mn$_4$-($S=8$), with an integer ground state S = 8. 
The third is the well 
known Fe$_8$ SMM with a $S = 10$ spin ground state~\cite{Sangregorio97}. 
The preparation, X-ray structure, and detailed 
physical characterization of both Mn$_4$ molecules have been presented 
~\cite{Yoo_Jae00,Bhaduri01}. The complex Mn$_4$-($S=9/2$) crystallizes in a hexagonal 
space group with a crystallographic C$_{\rm 3v}$ symmetry. The complex has a 
trigonal-pyramidal (highly distorted cubane-like) geometry. This 
complex is mixed-valent: Mn$_3^{\rm III}$Mn$^{\rm IV}$. The C$_{\rm 3v}$ axis 
passes through the Mn$^{\rm IV}$ ion and the triply bridging siloxide group. DC 
and AC magnetic susceptibility measurements indicate a well isolated 
$S = 9/2$ ground state~\cite{Bhaduri01}. 
The complex Mn$_4$-($S=8$) crystallizes in a 
triclinic lattice. The cation lies on an inversion center and 
consists of a planar Mn$_4$ rhombus that is also mixed-valent: 
Mn$_2^{\rm III}$Mn$_2^{\rm II}$. DC and AC magnetic susceptibility measurements 
indicate a $S = 8$ ground state~\cite{Yoo_Jae00}. 

All measurements were performed using an array of 
micro-SQUIDs~\cite{WW_PRL99}. The high sensitivity of this 
magnetometer allows us to study single crystals of SMMs
of sizes of the order of 10 to 500 $\mu$m.
The field can be applied in any direction by separately 
driving three orthogonal coils.

Before presenting our measurements we review briefly 
the giant spin model which is the simplest model describing the spin 
system of SMMs. 
The spin Hamiltonian is
\begin{equation}
	H = -D S_z^2 + H_{\rm trans} 
	+ g \mu_{\rm B} \mu_0 \vec{S}\cdot\vec{H} 
\label{eq_H}
\end{equation}
$S_x$, $S_y$, and $S_z$ are the three 
components of the spin operator; 
$D$ is the anisotropy constant defining an Ising
type of anisotropy; $H_{\rm trans}$, 
containing $S_x$ or $S_y$ spin operators, gives the transverse
anisotropy which is small compared to $D S_z^2$ in SMMs;
and the last term describes the Zeeman 
energy associated with an applied field $\vec{H}$. 
This Hamiltonian has an energy level 
spectrum with $(2S+1)$ values which, 
to a first approximation, can be labeled by the 
quantum numbers $m = -S, -(S-1), ...,  S$
taking the $z$-axis as the quantization axis. The energy spectrum
can be obtained by using 
standard diagonalization techniques.
At $\vec{H} = 0$, the levels $m = \pm S$ 
have the lowest energy. 
When a field $H_z$ is applied, the levels with 
$m < 0$ increase in energy, while those 
with $m > 0$ decrease. Therefore, 
energy levels of positive and negative 
quantum numbers cross at certain values of $H_z$. 
It turns out that the levels cross 
at fields given by $\mu_0 H_z \approx n D/g \mu_{\rm B}$, 
with $n = 0, 1, 2, 3, ...$. 

When the spin Hamiltonian contains transverse terms 
($H_{\rm trans}$), the level crossings  can be 
``avoided level crossings''. 
The spin $S$ is ``in resonance'' between two 
states when the local longitudinal field is close 
to an avoided level crossing. 
The energy gap, the so-called 
``tunnel spitting'' $\Delta$, can be tuned by 
a transverse field (a field applied perpendicular 
to the $S_z$ direction) 
via the $S_xH_x$ and $S_yH_y$ Zeeman terms.

The effect of these avoided level crossings can 
be seen in hysteresis loop measurements. 
Figs. 1a and 1b show typical hysteresis loop 
measurements for single crystals of Mn$_4$-($S=9/2$) 
and Mn$_4$-($S=8$) (for Fe$_8$ data, see~\cite{WW_PRL99}). 
When the applied field is near an 
avoided level crossing, the magnetization relaxes faster, 
yielding steps separated by plateaus. 
As the temperature is lowered, there is a decrease 
in the transition rate due to reduced thermally assisted tunneling. 
A similar behavior was observed in Mn$_{12}$ acetate clusters 
\cite{Novak95,Friedman96,Thomas96}. As seen for Fe$_8$ 
hysteresis loops become temperature-independent below 0.4 K
indicating ground state tunneling. The field between two resonances
allows us to estimate the anisotropy constants $D$. 
We obtain 0.68 K and 0.43 K for Mn$_4$-($S=9/2$) 
and Mn$_4$-($S=8$), respectively.

\begin{figure}
\begin{center}
\includegraphics[width=.42\textwidth]{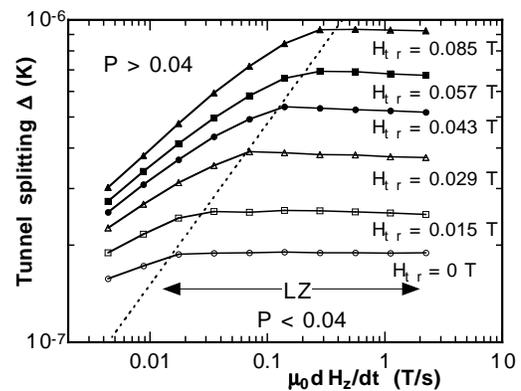}
\caption{Field sweeping rate dependence of the tunnel splitting 
$\Delta_{-9/2,9/2}$ 
measured by a Landau--Zener method for a Mn$_4$-(S=9/2) samples,
at different transverse fields. 
The Landau--Zener method
works in the region of high sweeping rates 
where $\Delta_{-9/2,9/2}$ is sweeping rate independent.}
\label{LZ_test}
\end{center}
\end{figure}

In order to establish the spin-parity effect, 
the tunnel splitting was measured as a function of transverse field. 
The field dependence of the tunnel splitting is expected
to be very sensitively dependent on
the spin-parity and the parity of the avoided level crossing. 
This approach is based on the 
Landau--Zener model that describes the nonadiabatic transition 
between the two states in a two-level system 
~\cite{Landau32,Zener32}. The original work by Zener concentrated on 
the electronic states of a diatomic molecule, while Landau 
considered two atoms that undergo a scattering process. Their 
solution of the time-dependent Schr$\ddot{\rm o}$dinger equation of a 
two-level system can be applied to many physical systems and it 
has become an important tool for studying tunneling transitions. The 
Landau--Zener model has also been applied to spin tunneling in 
nanoparticles and clusters 
~\cite{Miyashita95,Miyashita96,Rose98,Thorwart00,Leuenberger00}.

When sweeping the longitudinal field 
$H_z$ at a constant rate over an avoided energy level crossing,
the tunneling probability $P$ is given by

\begin{equation}
P_{m,m'} = 1 - {\rm exp}\left\lbrack-
     \frac {\pi \Delta_{m,m'}^2}
     {2 \hbar g \mu_{\rm B} |m - m'| \mu_0 dH_z/dt}\right\rbrack
\label{eq_LZ}
\end{equation}
Here, $m$ and $m'$ are the quantum numbers of the avoided level 
crossing, $dH_z/dt$ is the constant field sweep rate, 
$g~\approx~2$, $\mu_{\rm B}$ is the Bohr magneton, and $\hbar$ is 
Planck's constant. 

In order to apply quantitatively the Landau--Zener formula 
(Eq.~\ref{eq_LZ}), 
we first checked the predicted 
sweep field dependence of the tunneling rate. 
The SMM crystal was first placed in a high 
negative field to saturate the magnetization. 
Then, the applied field was swept at a constant rate 
over one of the resonance transitions and 
the fraction of molecules that 
reversed their spin was measured. The tunnel splitting $\Delta$ 
was calculated using Eq.~\ref{eq_LZ} and is plotted 
in Fig. 2 as a function of field sweep rate. The Landau--Zener method
is applicable in the region of high sweep rates 
where $\Delta_{-9/2,9/2}$ is independent of sweep rate .
The deviations 
at lower sweeping rates are mainly due to the 
{\it hole-digging mechanism}~\cite{WW_PRL99} 
as observed for Fe$_8$ SMM~\cite{WW_EPL00}.
Such behavior has recently been simulated~\cite{Liu01}
\footnote{
A more detailed study show that the tunnel
splittings obtained by the Landau--Zener method
are slightly influenced by environnemental effects
like hyperfine couplings~\cite{WW_EPL00}. Therefore, one might 
call it an effective tunnel splitting.
}.

\begin{figure}
\begin{center}
\includegraphics[width=.42\textwidth]{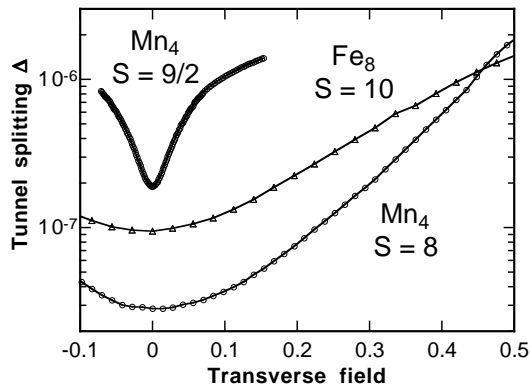}
\caption{Tunnel splitting for three SMMs
as a function of transverse field. The data for Fe$_8$ where
taken from ~\cite{WW_Science99} with the transverse field
applied along the medium hard axis.}
\label{delta}
\end{center}
\end{figure}

Fig. 3 presents the measured tunnel splittings obtained by 
the Landau--Zener method as a function of transverse
field. Note that for all three SMMs the tunnel splitting is
finite. However, depending on the spin-parity the sensitivity
to an applied transverse field is completely different.
The tunnel splitting increases gradually for an integer spin,
whereas it increases rapidly for a half-integer spin.
Note also that $\Delta$ is plotted on a logarithmic scale 
in Fig. 3, and
that the tunnel probability should depend on the
second power of $\Delta$ (Eq.~\ref{eq_LZ}). Therefore, our
measurements show that the tunneling rate 
of a half-integer spin is strongly transverse field dependent,
unlike the case for an integer spin SMM.  

\begin{figure}
\begin{center}
\includegraphics[width=.42\textwidth]{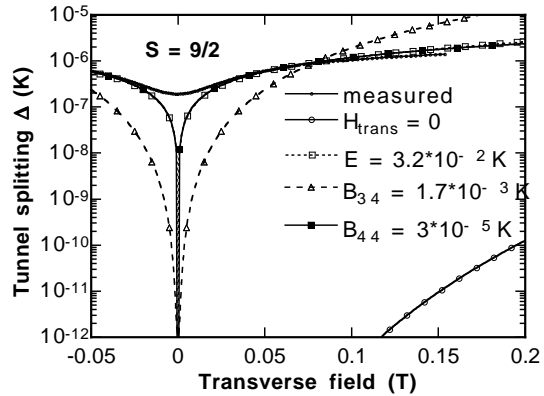}
\includegraphics[width=.42\textwidth]{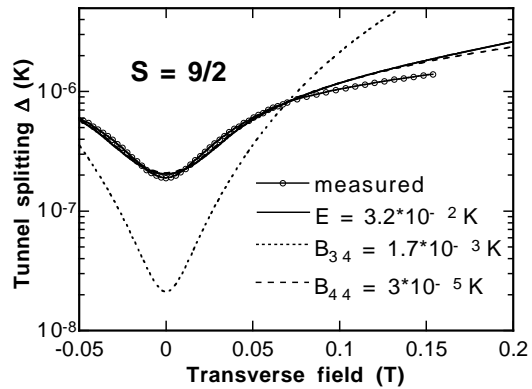}
\caption{(a) First attempt to simulated the 
measured tunnel splittings for Mn$_4$-($S=9/2$). 
For simplicity, the calculated $\Delta$ has been 
averaged over all possible orientation of the transverse field.
(b) Same as in (a) but taking into account a Gaussian 
distribution of transverse field components with a 
half-width $\sigma =$ 0.035 T.}
\label{delta_fit}
\end{center}
\end{figure}

Figs. 4a and 4b present the first attempt to simulate the 
measured tunnel splittings for Mn$_4$-($S=9/2$). 
Firstly, one should note that
there must be transverse terms in the spin Hamiltonian
because for $H_{\rm trans} = 0$, the measured tunnel splitting
should be  orders of magnitude smaller than observed. Secondly, we
tried to simulate the data with a transverse term 
($B_{34}/2*[S_z(S_+^3+S_-^3) + (S_+^3+S_-^3)S_z$) 
with respect to the C$_3$ symmetry
of the SMM. However, this term also cannot account for
the measured tunnel splitting. Finally, we found that either
the second order term ($E*(S_+^2+S_-^2)$) or a fourth
order term ($B_{44}*(S_+^4+S_-^4)$) can equally  well
describe the measurements~\footnote{
The fit could
be improved by  using a combination of transverse terms
but this does not lead to a better insight concerning
the spin parity effect.
}.
These results
suggest that  there is a small effect which breaks
the  C$_3$ symmetry. This could  be a small strain inside the
SMM crystal induced by defects. Such defects could result
from either a
loss of solvent molecules or 
dislocations~\cite{Chudnovsky01}.

In the ideal case, one  would expect  $\Delta = 0$ for $\vec{H} = 0$.
This is not found in a  real SMM because of the influence of 
environmental degrees of freedom that can induce tunneling. In our 
case, mainly hyperfine and dipolar couplings induce tunneling via 
transverse field components. Fig. 4b presents a simulation
of the  measured tunnel splitting when taking into account a Gaussian 
distribution of transverse field components with a width $\sigma =$ 
0.035 T. This value is in good
agreement with other SMMs.

The simulation of measured tunnel splittings for Mn$_4$-($S=8$)
is much easier because an integer spin is not very sensitive to 
transverse field components resulting from 
hyperfine and dipolar couplings. 
We found that a second order term with $E = 0.057$ K can describe well the 
data.

In conclusion, we have shown that the predicted spin parity effect 
~\cite{VanHemmen86,Loss92,Delft92} 
can indeed be observed by measuring the tunnel splitting as a function of 
transverse field. An integer spin system is rather insensitive to 
small transverse fields whereas a half-integer spin systems is
much more sensitive. However, a half-integer spin system will 
also undergo tunneling at zero external field as a result of 
environmental degrees of freedom such as hyperfine and dipolar couplings
or small intermolecular superexchange interaction.

I. Tupitsyn and M. Elhajal are acknowledged 
for fruitful discussions, and 
C. Thirion, D. Mailly and A. Benoit for 
support to the micro-SQUID technique.
This work was supported by the 
European Union TMR network MOLNANOMAG, 
HPRN-CT-1999-0012.

\end{document}